\documentclass[letter]{aa}
\usepackage[varg]{txfonts}

\usepackage{graphicx,rotating}
\usepackage{amssymb,amsmath,pdflscape}
\usepackage{multirow}
\usepackage{natbib}
\usepackage{upgreek}
\usepackage[colorlinks=true, linkcolor=blue, citecolor=blue, filecolor=blue, urlcolor=blue]{hyperref}

\begin{document}

\title{Equilibrium figure of Haumea and possible detection by stellar occultation}

\titlerunning{Equilibrium figure of Haumea and possible detection by stellar occultation}

\author{
    C. Staelen\inst{1}  
    \and N. Rambaux\inst{2} 
    \and F. Chambat\inst{3}
    \and J. C. Castillo-Rogez\inst{4}
}

\institute{
    Universit\'e Paris Cit\'e, CNRS, Institut de Physique du Globe de Paris, F-75005 Paris, France  \\ \email{staelen@ipgp.fr} 
    \and Sorbonne Université, Observatoire de Paris, Université PSL, CNRS, Laboratoire Temps Espace, F-75014 Paris, France
    \and Universit\'e de Lyon, CNRS, ENS Lyon, Laboratoire de Géologie de Lyon Terre - Planètes - Environnement, F-69007 Lyon, France
    \and Jet Propulsion Laboratory, California Institute of Technology, Pasadena, California, USA
}

\date{Received ??? / Accepted ???}

\abstract{
    The equilibrium figure of dwarf planet Haumea is studied to determine if the observed shape is compatible with a differentiated hydrostatic body. Three groups of interior models of Haumea are assumed, all with a rocky core and a volatile-rich outer shell that may contain some porosity. A third layer located between the core and the outer shell has a density suggesting partial differentiation or the presence of a large fraction of organic matter. Using the code BALEINES, which solves for the equilibrium figures of the boundaries between layers, we show that the hydrostatic models closest to the shape derived by stellar occultation approach a state of critical rotation, which translates into a pinched shape with large deviations from an ellipsoid (up to 110~km). The previous stellar occultation and light curves cannot distinguish between the ellipsoid and the pinched shape, but we predict this figure could be observable on the next stellar occultation of Haumea on May 4, 2026, if some chords are obtained in the northern or southern limbs of the shadow. 
}


\maketitle

\section{Introduction}

(136108) Haumea is a rapidly rotating Kuiper belt object whose triaxial figure was derived by \citet{ortiz17} from a stellar occultation together with light curves. However, \citet{kondra18} and \citet{ddp19} remarked this estimation assumed the occultation happened when Haumea's brightness was at a minimum, whereas a rotational phase of $-13.3^{\circ}$ was observed; this phase difference changes the set of ellipsoids that are compatible with the projected ellipse. In both cases, the resulting best-fit ellipsoids are incompatible with a Jacobi ellipsoid, i.e. a homogeneous ellipsoid in hydrostatic equilibrium (see Fig. \ref{fig:ba_ca}). This discrepancy implies that either rigid-body forces provide significant support against self-gravity, or Haumea has undergone differentiation and is thus heterogeneous. In the latter perspective, \citet{ddp19} and \citet{noviello22} have studied two layer hydrostatic models of Haumea, with a rocky core and an icy mantle. Their preferred model is nearly ellipsoidal and compatible with the chords of the 2017 occultation, meaning that Haumea can be consistent with differentiated hydrostatic bodies.  

A new stellar occultation by Haumea is predicted for May 4, 2026 \citep{ortiz26}\footnote{See also \url{https://lesia.obspm.fr/lucky-star/occ.php?p=156331}.}, which may refine the observational constraints on its shape. Therefore, this letter investigates triaxial equilibrium figures of two- and three-layer hydrostatic bodies to determine whether only nearly ellipsoidal shapes are likely to be observed. Unlike previous works, we show that, if Haumea has a hydrostatic shape, then it may deviate significantly from an ellipsoid and could even be pinched at the end of its equatorial major axis, indicating a state of critical rotation. Finally, we question whether observational data from May 2026 can discriminate an ellipsoid from a pinched figure.

\section{Hydrostatic shape models}\label{sec:methods}

The theory of equilibrium figures states that a rigidly rotating fluid mass is in hydrostatic equilibrium if and only if the surfaces of constant density and the surfaces of constant effective potential, i.e. the sum of the centrifugal and self-gravitational potentials, coincide \citep[e.g.][]{zt78,lan82}. Determining an equilibrium figure then comes down to finding shapes of the fluid whose effective potential is constant on these surfaces. As the potential depends non-trivially on the shape, the corresponding equations set actually forms a fixed-point problem. In this paper, we consider a differentiated body composed of $L=2$ or $3$ homogeneous layers, indexed by $\ell$  with $\ell=1$ denoting the innermost layer and $\ell=L$ the outermost. Assuming homogenous layers is sufficient since under the low pressures inside Haumea ($\lesssim500$ MPa), the densities of ices and rocky minerals increase only by a few percent (the bulk moduli are $\sim10$ GPa for ices and $\gtrsim 100$ GPa for rocks). 

We use the code BALEINES \citep{sh26}, which computes iteratively the shapes of the interfaces. Each surface is expanded over Chebyshev polynomials both for latitude and longitude, the potential is written in term of surface integrals only, and these integrals are computed with Clenshaw-Curtis quadratures. Mirror symmetry with respect to the plane $(x{\rm O}y)$ is imposed by the Lichtenstein theorem, and mirror symmetry with respect to the planes $(y{\rm O}z)$ and $(z{\rm O}x)$ is assumed to reduce the computation times, $\rm O$~being the centre of mass, $({\rm O}x)$, $({\rm O}y)$ and $({\rm O}z)$ coinciding with its equatorial major, equatorial minor and polar axes, respectively. Here, $17$ Chebyshev nodes are used both for latitude and longitude, which results in a relative precision of the order of $10^{-4}$. 

Let $a_{\ell}$, $b_{\ell}$ and $c_{\ell}$ be the equatorial major semi-axis, the equatorial minor semi-axis and the polar axis of the boundary of layer~$\ell$ and $\rho_{\ell}$ its mass density. The explored parameters of the models reported in this work are $c_L/a_L$, $\rho_{\ell}/\rho_{\ell+1}$ ($1\leq\ell<L$) and $c_{\ell}/c_L$. 
Starting from two- and three-layer models of Haumea, we explore a large range of this parameter space: $\rho_{\ell}/\rho_{\ell+1}\in[1.2,4.2]$, $c_1/c_2 \in [0.25,0.95]$ for two layers, $c_2/c_3\in[0.30,0.95]$ and $c_1/c_3\in{[0.25,c_2/c_3[}$ for three layers, and $c_L/a_L\in[0.42,0.52]$. The results are summarised in Fig.~\ref{fig:ba_ca}, which displays, in the $(c_L/a_L,b_L/a_L)$-plane, the region where hydrostatic models have been found. As expected, heterogeneity expands the possible ranges of $b_L/a_L$ for a given $c_L/a_L$, while only a unique value is permitted in the homogeneous case. The hydrostatic figures are contained between two curves into the $(c_L/a_L,b_L/a_L)$-plane: the curve at higher $c_L/a_L$ is given by models with a large outer layer \citep{sh26}, while the curve at lower $c_L/a_L$ corresponds to configurations at the mass-shedding limit, that is to say the centrifugal acceleration, $\varOmega^2a_L$, $\varOmega$ being the rotation rate, exactly balances the self-gravity at the end of the equatorial major axis, $g_a$. To measure the proximity to this limit, we define $q=\varOmega^2a_L/|g_{a}|$, so that the mass-shedding limit, or critical rotation, is defined by $q=1$. 

\section{Compatibility with the observational data}\label{ssec:hydroshape}

\begin{table*}[ht]
    \centering
    \caption{Examples of hydrostatic solutions whose free surface is coherent with the chords obtained during the 2017 occultation \citep{ortiz17}.}
    \begin{tabular}{lcccccccccccccccc}
        \hline
        &$L$ & \multirow{2}{*}{$\dfrac{c_L}{a_L}$} & \multirow{2}{*}{$\dfrac{\rho_1}{\rho_2}$} & \multirow{2}{*}{$\dfrac{\rho_2}{\rho_3}$} & \multirow{2}{*}{$\dfrac{c_1}{c_L}$} &\multirow{2}{*}{$\dfrac{c_2}{c_L}$} & $q$ & $a_L$ & $c_L$& $\bar{\rho}$ & $\rho_3$ & $\rho_2$ & $\rho_1$ & $\tau$ & $J_2$ \\
        & & & & & & & & $({\rm km})$ & $({\rm km})$ & $({\rm kg/m^3})$ & $({\rm kg/m^3})$ & $({\rm kg/m^3})$ & $({\rm kg/m^3})$ & $({\rm hr})$ & \\\hline
        A & $3$ & $0.513$ & $1.4$ & $2.6$ & $0.45$ & $0.90$ & $0.59$ & $1046$ & $537$ & $2055$ & $936$ & $2435$ & $3408$ & $3.92$ & $0.158$  \\
        B & $3$ & $0.450$ & $1.7$ & $2.3$ & $0.66$ & $0.90$ & $0.97$ & $1227$ & $552$ & $1932$ & $858$ & $1974$ & $3356$ & $3.91$ & $0.157$ \\
        C & $3$ & $0.460$ & $2.6$ & $1.4$ & $0.78$ & $0.90$ & $0.99$ & $1170$ & $538$ & $2078$ & $952$ & $1333$ & $3466$ & $3.67$& $0.140$\\
        D & $3$ & $0.487$ & $1.8$ & $2.6$ & $0.38$ & $0.86$ & $0.73$ & $1126$ & $547$ & $1952$ & $972$ & $2528$ & $4551$ & $3.92$ & $0.159$ \\
        E & $3$ & $0.496$ & $2.2$ & $1.4$ & $0.80$ & $0.95$ & $0.68$ & $1095$ & $543$ & $1984$ & $938$ & $1313$ & $2890$ & $3.91$ & $0.158$ \\
        F & $2$ & $0.460$ & $3.8$ & - & $0.82$ & - & $0.98$ & $1163$ & $536$ & $2081$ & - & $880$ & $3346$ & $3.65$ & $0.140$\\\hline
    \end{tabular}
    \label{tab:results}
    \tablefoot{In all configurations, $M=3.952\times10^{21}~{\rm kg}$ and $b_L=840~{\rm km}$. The reference radius for $J_2$ is 798~km as that of \cite{proudfoot24}.}
\end{table*}

All quantities in BALEINES are normalised by the length of the equatorial major semi-axis and the mass density of the outermost layer of the fluid mass, $a_L$ and $\rho_L$. In order to compare the outputs with data, we must come back to dimensional quantities. These physical outputs are obtained by requiring that the mass matches the value currently determined for Haumea, i.e. $M=3.952\times10^{21}~{\rm kg}$ \citep{proudfoot24}, and that the length of the minor equatorial semi-axis $b_L=840~{\rm km}$ \citep{ddp19}. We impose $b_L$ since it is the best constrained semi-axis length, with estimated values between $830\pm6~{\rm km}$ \citep{kondra18} and $852\pm4~{\rm km}$ \citep{ortiz17}. Tests with both these lengths have shown no significant difference with the value chosen here.

\begin{figure}[t]
    \centering 
    \includegraphics[width=0.9\linewidth,trim={0 0.2cm 0 0},clip]{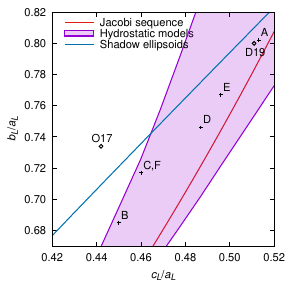}
    \caption{Possible hydrostatic models obtained in this work (pink region). The models of Table~\ref{tab:results} are plotted with black crosses. For reference, the Jacobi sequence is plotted in red line. The best-fit ellipsoids of \citet{ortiz17} (``O17'') and the hydrostatic model preferred by \citet{ddp19} (``D19'') are reported with open circles. The ``shadow ellipsoids'' curve corresponds to ellipsoids whose projection corresponds to the ellipse determined during the 2017 occultation, with the pole orientation obtained by \citet{ortiz17} and a rotational phase of $-13.3^{\circ}$; see \citet{drummond1985} for the equations used. Note that this curve is valid only for ellipsoidal shapes.} 
    \label{fig:ba_ca}
\end{figure}

\begin{figure}[ht]
    \centering
    \includegraphics[width=0.95\linewidth,trim={0 0.5cm 0 0.2cm},clip]{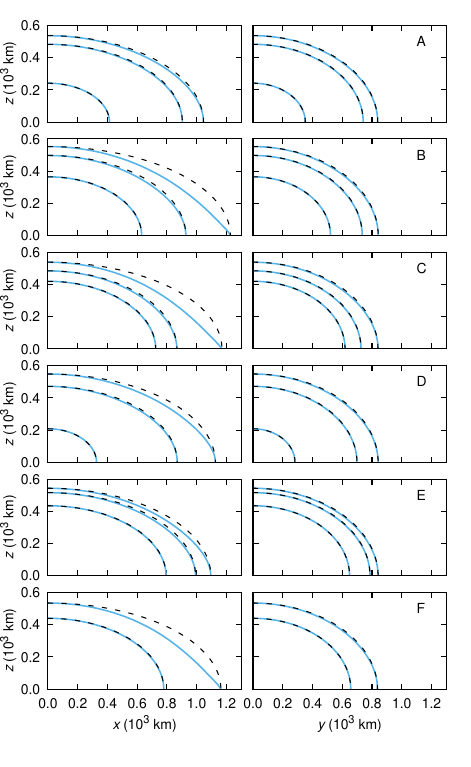}
    \caption{Hydrostatic shape models for Haumea in the $(x{\rm O}z)$ (left) and $(y{\rm O}z)$ (right) planes, corresponding to the configurations reported in Table \ref{tab:results}. The dashed lines correspond to the ellipsoids with the same axis lengths as the layers' boundary.}
    \label{fig:results}
\end{figure}

From the explored space, we select six configurations, labelled from A to F,  whose free surface shapes are coherent with the chords obtained during the 2017 occultation event. Fig.~\ref{fig:results} shows that their boundaries are quite elliptical in the $(y{\rm O}z)$ plane but significantly deviate from an ellipse in the $(x{\rm O}z)$ plane, up to pinched shape for configurations B, C and F. The maximal deviations from an ellipsoid are approximatively of $110~{\rm km}$ for the pinched configurations. The properties of these six configurations are reported in Table \ref{tab:results} and in Fig. \ref{fig:ba_ca}. 

\begin{figure}[ht]
    \centering
    \includegraphics[width=0.9\linewidth,trim={0 0 0 0.2cm},clip]{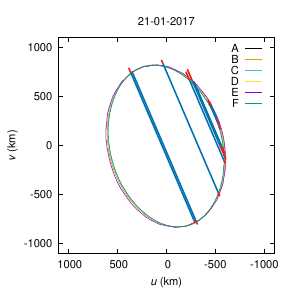}
    \caption{Projections of the six hydrostatic representative solutions in the sky plane at the epoch of the 2017 occultation. $u$ and $v$ are the celestial East and North, respectively. The pole orientation is the one obtained by \citet{ortiz17} and the rotational phase is $-13.3^{\circ}$.}
    \label{fig:2017_occ}
\end{figure}

We also project their free surface onto the sky plane at the epoch of the 2017 occultation; see Fig.~\ref{fig:2017_occ}. To do so, we use the Chebyshev expansion to resample the shape of the free surface, and form a regular triangular mesh with 15155 facets that we use as an input file for the open-access tool developed by \citet{rizos25}. These projections confirm that hydrostatic models are able to fit the occultation chords obtained by \citet{ortiz17}. Light curves obtained with these shapes are reported in Appendix \ref{app:lightcurves} and show that hydrostatic models are also able to fit the light curves obtained by \citet{lacerda2008}, \citet{lockwood14} and \citet{ortiz17}. Finally, the $J_2$ values of these configurations are compatible with the corresponding observation: they range between $0.140$ and $0.159$, which lies near the bottom of the $1\upsigma$ confidence region obtained by \citet{proudfoot24} by orbit fitting, i.e. $J_2^{\rm obs} = 0.262^{+0.103}_{-0.112}$.

\section{Discussion}

\subsection{Haumea as a critical rotator}

Some hydrostatic figures exhibit a pinched geometry, i.e. the surface normal becomes discontinuous at the end of the major equatorial axis. Since the effective gravity is continuous and, for a hydrostatic body, orthogonal to the equipotential surfaces, this behaviour is possible only if the effective gravity vanishes at that point, i.e. for critical rotators ($q=1$). This is illustrated by configurations B, C and E, whose values of $q$, reported in table \ref{tab:results} are between $0.96$ and $0.99$. It is noteworthy that pinched shapes for critical rotators are actually well documented in the context of stellar physics \citep[e.g.][]{maeder12}. Furthermore, such equatorial ridges have been observed at several fast-rotating asteroids as (2867) \v{S}teins \citep{jorda12} and (65803) Didymos \citep{barnouin24}. Also, note that even configurations D are close to the mass-shedding limit ($q=0.73$, which corresponds to $85~\%$ of the critical rotation rate). Critical rotators are closer to the observed rotation period as $c/a$ approaches $0.45$ (configuration B), while for $c/a \gtrsim 0.49$ only figures close to ellipsoids are compatible to the rotation period (configurations A and E).

\subsection{Haumea's shape as a fossil figure}

As shown by configurations A, B and D, some models are fully compatible with the 2017 occultation and the observed rotation period of Haumea. However, Haumea's shape could also be ``frozen'', i.e. representative from a previous rotation state, and would then depart from hydrostatic equilibrium; this is why we have not imposed our models to exactly fit the observed rotation period. The rotation periods of configurations C and F are $14.7$ to $15.6$ minutes shorter than the observed value, implying a relative difference of roughly $6.5~\%$. In the formation scenario hypothesized by \citet{noviello22}, Haumea's rotation was slowed through hydration of the rocky core by hydrothermal circulation that  generated a redistribution of mass and increased its moment of inertia. Tidal migration of satellites represents another mechanism capable of decreasing a planet's spin rate. 

From the conservation of angular momentum in the whole system, and assuming circular orbit in the equatorial plane of Haumea for the satellites, one can derive the coupled evolution of the primary's rotation and the satellites' orbits \citep{md99}. However, considering the low masses and the orbital parameters of Hi'iaka and Namaka obtained by \citet{proudfoot24} and $k_2/Q \lesssim 10^{-2}$, the computed timescales are longer than the age of the Solar system. Thus, tidal migration alone cannot account for Haumea's present spin state. Additional spin-down could have occurred through small impacts, potentially involving debris from its ring. However, as demonstrated by configuration B, critical rotators can also be consistent with the observed rotation period, implying that a spin-down is not mandatory for Haumea to have a pinched shape.

\subsection{Internal structures}
The mean densities of our hydrostatic models range from $1932$ to $2081~{\rm kg/m^3}$, which lies between Pluto's \citep[$1853\pm4~{\rm kg/m^3}$,][]{brozovic24} and Eris' \citep[$2430\pm50~{\rm kg/m^3}$,][]{holler21}, and is consistent with the densities expected for material condensed in the Kuiper belt \citep{barr2016interpreting}. 

Composition-wise, three groups of models have been identified from our exploration of the parameter space. All three of them can be summed up as a rocky core and a thick volatile-rich shell, coherent with the structures reported by \citet{ddp19} and \citet{noviello22}. The first one consists in a two-layer structure with a ``low'' mass density shell, up to $830~{\rm kg/m^3}$ thick overlaying a rocky core, like configuration F in table \ref{tab:results}. The low shell density may be interpreted as 12-17\% of porosity (averaged), assuming a pure ice composition, or about 55\% porosity assuming an undifferentiated water-rock-organic-rich shell and taking as a reference the aforementioned mean density of  $1900~{\rm kg/m^3}$. 
The second group consists in three-layer configurations, among which configurations A, B, C and E, with an outer volatile-rich shell, a rocky core, and a thin mantle in between. The latter has a density between $1200$ and $1500~{\rm kg/m^3}$. This may indicate partial separation of the rock from the volatile phase or the presence of a high fraction or insoluble organic matter \citep{mckinnon2020formation}. In this line of models, the shell could have a density as high as $980~{\rm kg/m^3}$, corresponding to a non-porous, pure ice region. 
The last group of models also consists in three-layer structures, including configurations D, similar to the the two-layer models but with the addition of a small inner core with a density $\geq 4500~{\rm kg/m^3}$, which suggests the concentration of metal-rich phases (e.g., magnetite) mixed with silicates.
    
From a thermal evolution standpoint, models of Uranus' moon Titania are relevant since the two objects share similar physical properties (density, size).  \citet{castillo2023compositions} found that Titania could preserve a thick ($\sim 80$ km) outer layer that is a porous mixture of ice and rock. Since Haumea's surface is colder than Titania's by 20 to 50 K, one should also expect the dwarf planet to also preserve an undifferentiated outer layer \citep[see also][]{desch09}, a case that is consistent with the shell density derived from the shape. Note that the models of \citet{noviello22} do not include an undifferentiated crust, as they assume Haumea formed in a collision between two large Kuiper-belt objects that removed both their crusts. This scenario supposes a high-velocity impact; however, a lower-energy collision could plausibly allow the crust to survive.


\begin{figure}[t]
    \centering
\includegraphics[width=0.95\linewidth]{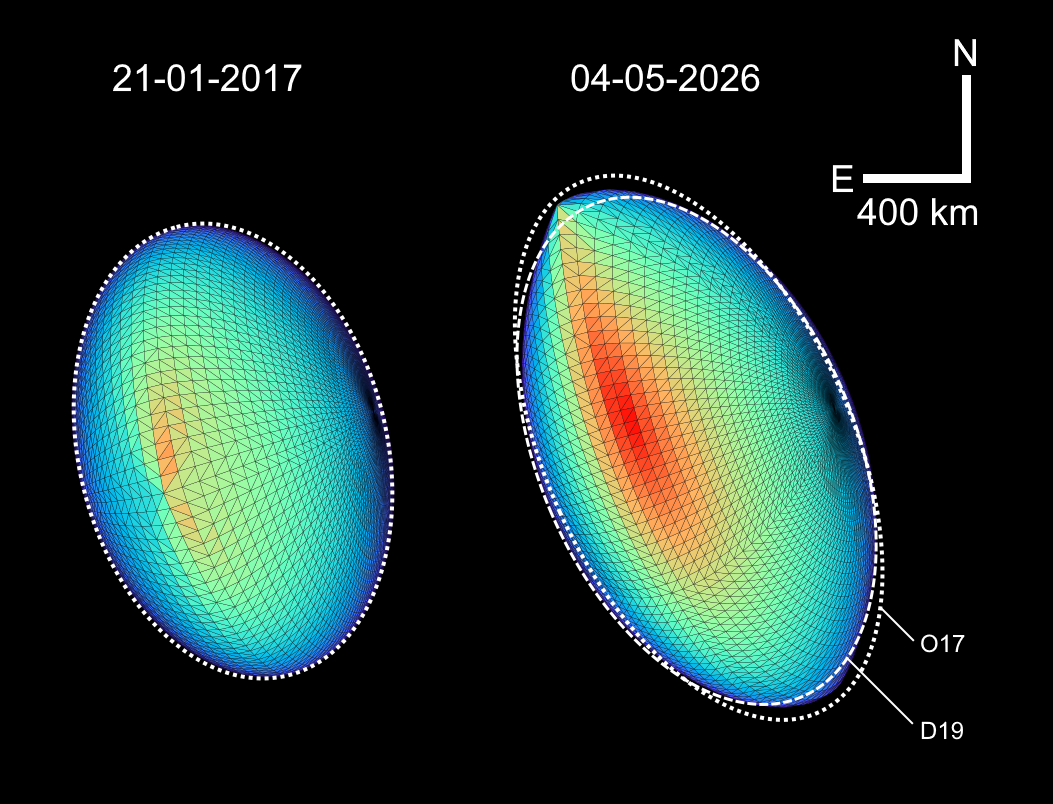}
    \caption{Projections of the hydrostatic shape of configuration C in the sky plane at the time of the 2017 occultation (left) and the future 2026 occultation (right). The projected ellipse obtained by \citet{ortiz17} and the projection of the ellipsoids of \citet{ortiz17} (O17) and \citet{ddp19} (D19) are shown on the left and right panels, respectively, in white dotted line for comparison.  The facet colours code the relative irradiance they receive, red being the highest and purple-blue the lowest.}
    \label{fig:2026_occ}
\end{figure}

\subsection{Expectations for the 2026 occultation}\label{ssec:projection}

As mentioned above, Haumea's equatorial major axis was not visible at the time of the 2017 occultation, making the possible deviation to the ellipsoid invisible. However, at the time of the predicted occultation of May 4, 2026, the rotational phase would be $0.92$, i.e. $\sim 25^{\circ}$ from the brightness maximum \citep{ortiz26}, meaning the equatorial major axis could be visible. As a consequence, the projected shape could greatly deviate from an ellipse. Fig.~\ref{fig:2026_occ} shows the predicted shape for configuration C, which corresponds to the case where the deviation between the hydrostatic model and the ellipsoids of \citet{ortiz17} and \citet{ddp19} is the most visible, and thus the most likely to be detected by stellar occultation. If Haumea is very close to the critical rotation, we then expect a pinched shape that would be visible at the ends of the equatorial major axis. Otherwise, we expect the shape to be closer to an ellipse. Thus, the most decisive parts to observe in order to distinguish between the models, and thus conclude on the hydrostaticity of Haumea, are the northern and southern limbs of the shadow. Should the pinched shape be observed with this stellar occultation, this would support the assumption that Haumea is close to a hydrostatic body at the critical rotation and thus would support the scenario of the creation of the haumean family by mass shedding proposed by \citet{ortiz12} and \citet{noviello22}.

\begin{acknowledgements}
    We are grateful to B. Sicardy for fruitful discussions about the occultations of 2017 and 2026 and to S. Charnoz for informed comments on the manuscript before submission. We thank the referee, S. Desch, for pointing out the problem with the rotational phase in Ortiz et al. (2017). This work has been supported by the French Agence Nationale de la Recherche, project Roche, number ANR-23-CE49-0012.
\end{acknowledgements}

\bibliographystyle{aa}
\bibliography{staelen_haumea}

\begin{appendix}

\section{Light curves}\label{app:lightcurves}

\begin{figure}[ht]
    \centering
    \includegraphics[width=0.9\linewidth]{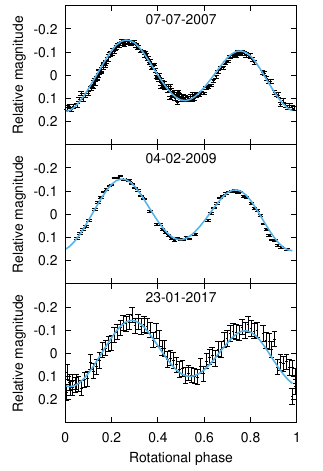}
    \caption{Comparisons between the light curves of Haumea reported by \citet[][top]{lacerda2008}, \citet[][middle]{lockwood14} and \citet[][bottom]{ortiz17} and synthetic light curves obtained with the hydrostatic figures.}
    \label{fig:lightcurves}
\end{figure}

The Hapke model was used for the reflectance of Haumea, with the same parameters as \citet{lockwood14} and a darker spot covering a quarter of Haumea with a 10~\% albedo variation \citep{lacerda2008}. We reproduce three light curves reported in \citet{lacerda2008}, \citet{lockwood14} and \citet{ortiz17}; the results are shown in Fig.~\ref{fig:lightcurves}. Note that only the light curve obtained with configuration C is plotted in Fig.~\ref{fig:lightcurves}, but the curves obtained with the other configurations are actually similar to the one reported here. The pinched shape leaves a mark in the light curve, with slightly narrower maxima and slightly larger minima. However, this feature would be difficult to detect on real light curves due to the limited resolution or albedo variations on the surface. Thus, it appears that only stellar occultation could allow us to observe if Haumea has an ellipsoidal or a pinched shape.

\end{appendix}

\end{document}